\documentclass[a4paper]{article}
\usepackage{graphicx}
\usepackage{subcaption}
\usepackage[titletoc]{appendix}
\usepackage{fullpage}
\usepackage{textcomp,gensymb}
\usepackage{amsmath}
\usepackage{cite}
\usepackage[english]{babel}
\usepackage[none]{hyphenat}
\usepackage{parskip}
\usepackage{amsfonts}
\usepackage{caption}
\usepackage{float}
\usepackage{cleveref}
\usepackage{multicol}
\usepackage[separate-uncertainty = true]{siunitx} 
\usepackage{url}
\usepackage{titling}
\usepackage{authblk}

\usepackage[colorinlistoftodos]{todonotes}

\usepackage{chapterbib}

\usepackage{enumitem}

\usepackage[all]{nowidow}

\usepackage[raggedright]{titlesec}

\usepackage{tcolorbox}

\usepackage{soul}
\setuldepth{\textbf{Wouter}}

\usepackage{psfrag}
\usepackage{overpic}
\usepackage{xcolor}

\usepackage{miller} 

\newcommand{\PbSnTe}{Pb$_{1-x}$Sn$_x$Te }
\newcommand{\PbSnTehalf}{Pb$_{0.5}$Sn$_{0.5}$Te }
\DeclareUnicodeCharacter{2212}{-}
\DeclareUnicodeCharacter{2212}{-}
\newcommand{\RNum}[1]{\uppercase\expandafter{\romannumeral #1\relax}}

\begin{document}
\title{Understanding the anisotropic growth of VS grown \PbSnTe nanowires}
\author[1]{Mathijs G.C. Mientjes}
\author[1]{Xin Guan}
\author[1,2]{Marcel A. Verheijen}
\author[1]{Erik P.A.M. Bakkers}
\affil[1]{Eindhoven University of Technology, 5600 MB Eindhoven, The Netherlands}
\affil[2]{Eurofins Materials Science Eindhoven, 5656 AE Eindhoven, The Netherlands}
\newpage
\begin{titlingpage}
    \maketitle
    \begin{abstract}
        \PbSnTehalf is a topological crystalline insulator (TCI), which holds promise for scattering-free transport channels and fault-tolerant quantum computing. As the topologically non-trivial states live on the surface, the nanowire geometry, with a high surface-to-volume ratio, is ideal for probing these states. The controlled growth of \PbSnTehalf nanowires using molecular beam epitaxy has been shown before, but an understanding of the anisotropic growth and the resulting morphology is lacking. Here, based on experimental observations, we develop a model that describes the evolution of NW morphology as a function of growth time. It is found that the anisotropic morphology can be described by a combination of direct impingement, mask diffusion and facet diffusion which results in a transition from a Te-limited growth regime to a group IV-limited growth regime. This growth model allows us to design more targeted experiments which could lead to a higher flexibility in device design. 
    \end{abstract}
\end{titlingpage}

\section{Introduction}
Topological Crystalline Insulators (TCIs) are a relatively new material class that holds promise for topological quantum computation as well as application in spintronic devices due to their topologically protected, spin-polarized surface states \cite{Fu_2006_Surface_states} \cite{Fu_2007_top_insulator_} \cite{Schindler_2018_HOTI}. One such material is \PbSnTehalf \cite{Hsieh_2012_SnTe} \cite{Dimmock_1966}. As the topological states live on the surface of the material, nanowires, which have a high surface-to-volume ratio, are an ideal geometry. As shown in our previous work, we can grow single crystalline nanowires with the rock salt crystal structure in regular arrays with a tunable aspect ratio ranging from 3-20 using a vapor-solid (VS) growth mechanism by Molecular Beam Epitaxy (MBE)\cite{mientjes2024catalyst}. The high aspect ratio of MBE grown nanowires is usually induced by the presence of a catalyst particle \cite{wagner1964vapor} \cite{wacaser2009preferential}, which increases the axial growth rate, or from intrinsically different growth rates on facets from different crystal families.\cite{hsu2005vapor} \cite{dimakis2011self}. 
However, the \PbSnTehalf nanowires are grown by a VS growth mechanism and are terminated by facets of the \{100\} family and all facets have thus identical surface energies\cite{mientjes2024catalyst}. Therefore, these relatively high aspect ratios must be caused by other growth factors. Here we aim to elucidate the relevant growth factors and determine their weight by modeling these growth factors and comparing this model to the evolution of morphology of MBE-grown nanowires with time. A better understanding of the underlying growth mechanism could lead to a higher tunability of NW morphology.

\begin{figure}[H]
\centering
\includegraphics[width=1\textwidth]{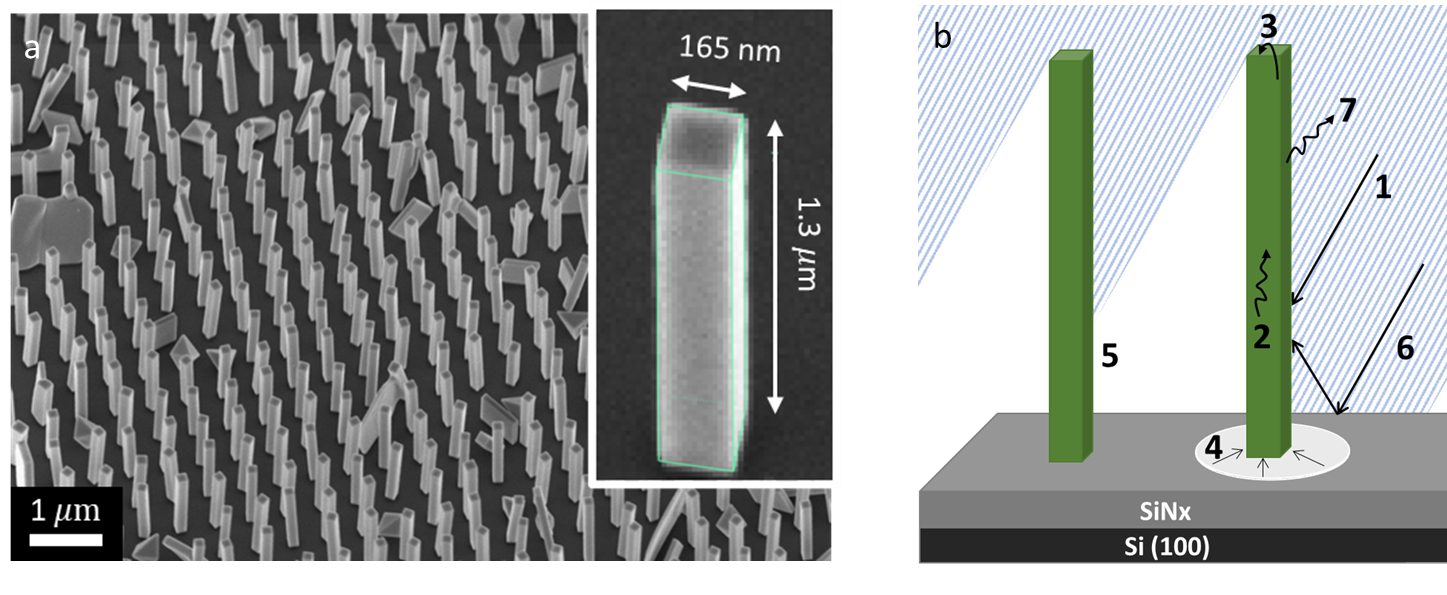}
\caption{ \fontsize{8}{14}\selectfont \textbf{Growth of \PbSnTehalf NW arrays and MBE growth factors} (a) SEM image of a representative NW array of \PbSnTehalf nanowires with a pitch of 500 nm and a hole size of 30 nm. In the inset, typical dimensions of such a NW are shown. b) A schematic visualizing the possible growth factors for VS-grown NWs discussed in this work. 1: Direct impingement, 2: adatom diffusion along the NW facet, 3: adatom diffusion across facets, 4: adatom diffusion on the mask, 5: shadowing, 6: reflection from the mask, 7: evaporation from the NW.}
\end{figure}

\section{Methods}
 The NWs in this work are grown using a similar method as reported in REF \cite{mientjes2024catalyst}. The NWs grow from holes in a 20 nm thick SiNx mask on a [100]-oriented Si substrate in ultra-high vacuum at a substrate temperature of 339 \degree C. The sample consists of multiple fields containing a square pattern of holes with varying hole sizes (20 nm-100 nm) and pitch (500 nm-2000 nm). Material is deposited from Pb and Sn effusion cells and a Te cracker cell under a 30-degree incident angle with respect to the substrate normal. All sources are located on the same side of the chamber, where there is a 90-degree angle between the Sn and Te source and a 45-degree angle between the Te and the Pb source (See SI 1). All presented samples are grown with a Group IV to Group VI ratio of the fluxes of 1/3 and a Pb-to-Sn ratio of 1 ($x_{input}=0.5$) measured using a beam flux monitor. During the growth, the sample is kept rotating at 5.3 RPM. A more in-depth discussion of the fabrication can be found in SI 2.

\section{Results}
\subsection{Nanowire morphology}

The NW morphology for each field has been studied using Scanning Electron Microscopy (SEM). The result of a typical growth run can be seen in Figure 1a. The NWs show good yield (typically around 50\%) and dimensional uniformity allowing for a quantitative analysis of their morphology. These wires, for example, have an average width and length of 165 nm and 1300 nm, respectively (see inset). This gives an average aspect ratio of 7.8, which is much larger than the naively expected aspect ratio of 1 for a cubic crystal terminated by 6 symmetrically equivalent \{100\} facets. An automated SEM image analysis script has been applied to quantitatively determine the dimensions, (length, width, volume, and aspect ratio) \cite{APIMs}. Using this method, the morphology of hundreds of nanowires per field has been investigated, allowing for a statistically meaningful analysis. 

Using this method, we studied the evolution of the average NW morphology with time. In Figure 2, a sub-set of the data is shown (Markers) for a pitch of 1 $\mu$m and hole sizes ranging from 20 to 100 nm. The nanowire length initially increases linearly in time and then levels off (figure 2a), whereas the nanowire width (figure 2b) constantly increases in time. These together lead to a super-linear increase in NW volume (Figure 2c) with growth time. Due to the decrease in axial growth and the constant radial growth rate, the NW aspect ratio (Figure 2d) decreases with growth time. This data set is representative for all experiments and is used to develop the growth model. In order to understand the observed trends, we first identify the relevant growth factors.

\begin{figure}[h]
\centering
\includegraphics[width=1\textwidth]{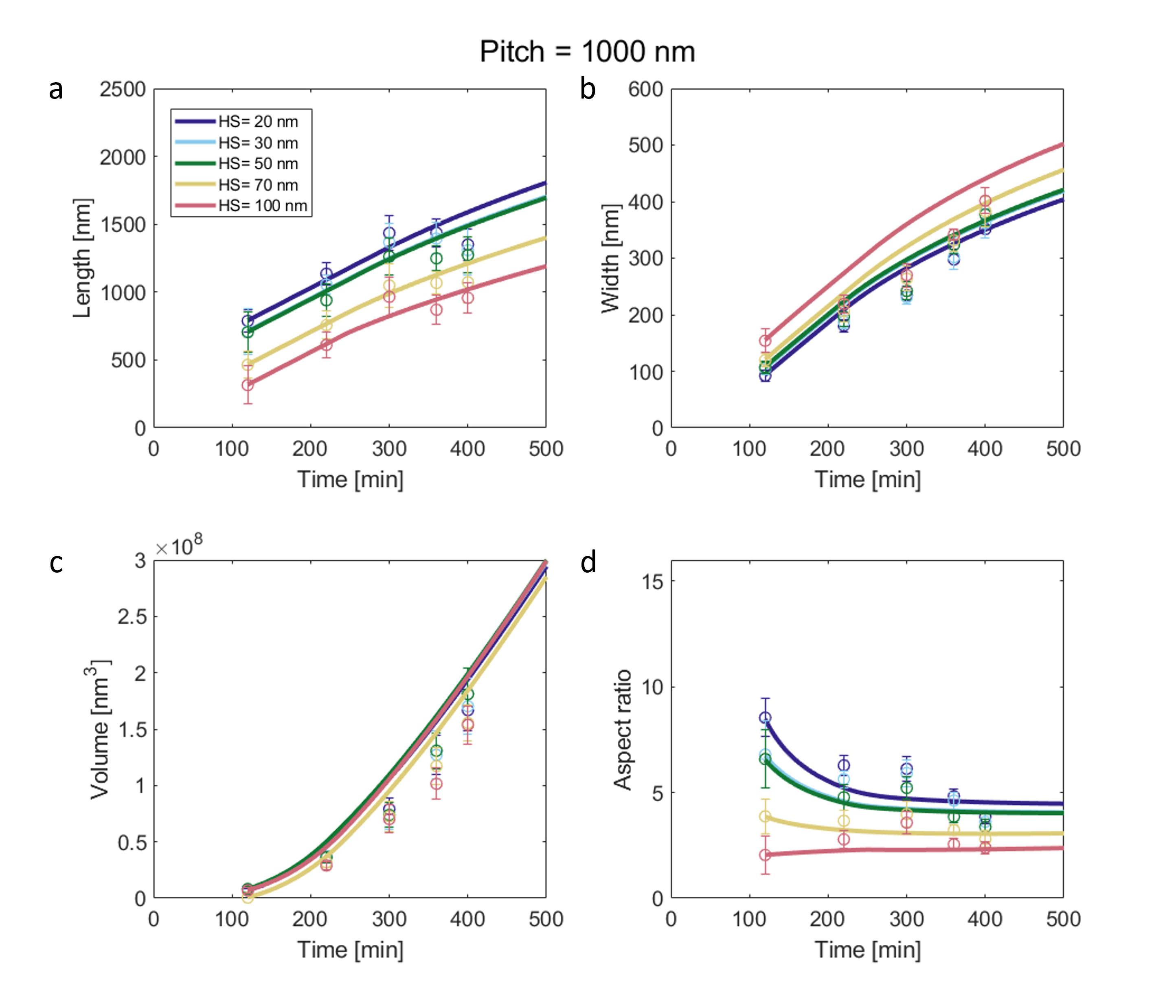}
\caption{ \fontsize{8}{14}\selectfont \textbf{NW morphology} a-d) The average NW length, width, volume and aspect ratio evolution with time (markers) for a pitch of 1000 nm and multiple hole sizes, respectively. The solid lines are the prediction of the growth model of the NW morphology using the growth factors: Direct impingement, shadowing, mask diffusion and facet-to-facet transfer with $\lambda_{mask, Sn}$=350 nm, $\lambda_{mask, Pb}$=300 nm and $\lambda_{mask, IV-Te}$=250 nm.}
\end{figure}

\subsection{Relevant growth factors}
Figure 1 b schematically visualizes all possible growth factors for VS MBE growth of nanowires. In order to build a numerical model for the growth of \PbSnTehalf NWs, which will be presented below, the relevant factors need to be quantified. Below, we individually present all growth factors and discuss their growth contributions by defining the input parameters for the growth model. 

\begin{enumerate}
    \item \textbf{Direct impingement}. Due to the ultra-high vacuum and the chamber dimensions, the incoming materials can be considered as parallel rays of material \cite{Dubrovskii_2022_Adsorbing}. The flux from direct impingement arriving on the different NW facets depends on the total amount of flux and the angle of the beam with respect to the NW facets. 
    \\As input for our model, the absolute fluxes of atoms of the different species have been determined by measuring the growth rate of a planar \PbSnTehalf layer, grown under identical conditions. This growth rate was found to be 1.2 $\pm$ 0.2 nm/min (supplementary). Since \PbSnTe grows stoichiometrically and the growth is limited by the Group IV supply (from the IV/VI ratio), we assume the total Group IV flux to be half the planar growth rate, $F_{IV}=0.575$ nm/min. Since the IV/VI ratio is 1/3, the Te flux is estimated to be three times the Group IV flux, $F_{Te}=1.725 $ nm/min. We assume that the excess Te evaporates due to the high vapor pressure \cite{Springholz_2002}. 
    \\ In the case of modeling a NW during growth, different effective fluxes are implemented for top and side facets based on their respective angle with the incoming material. Furthermore, as the sample rotates during the growth, the nanowire top facet is constantly exposed whereas the side facets are only exposed a fraction of the time. This leads to lower effective flux on the side facets compared to the top facet.

    \item \textbf{Diffusion on the NW facets.} Adatoms on NW facets can diffuse  \cite{jensen2004role}, and have a species-specific surface diffusion length, $\lambda_{NW}$. We consider Pb, Sn and Te to be present on the NW facets. If diffusion lengths are shorter than the length of the wire and material distribution is inhomogeneous over the side facets (e.g. by shadowing or from mask diffusion), tapering will occur\cite{harmand2005analysis}. In the current data set, no tapering of the nanowires is observed for nanowires up to a length of 2 microns, as can be seen in Figure 1a, indicating that diffusion lengths are longer than the NW dimensions or dwelling times are very short. From a series of experiments (See SI 3.2), we know that Pb and Sn have a long dwelling time on the NW facets and from the non-tapered NW geometry we conclude that $\lambda_{NW,Sn}$, and  $\lambda_{NW,Pb}$ are longer than the length of the NWs. 
    It is known that Te has a vapor pressure orders of magnitude higher than that of the other species \cite{Springholz_2002}, such that Te will quickly evaporate, leading to a short dwelling time and thus a short diffusion length, $\lambda_{NW, Te}$, on the NW facets.

    \item \textbf{Facet to facet transfer.}
    Experiments, in which the sample is not rotated, show there is growth on the unexposed facets, indicating that adatoms can diffuse from one facet to another (see SI 3.3). This facet-to-facet diffusion may be facilitated by the rounding of the corners at the atomic scale \cite{mientjes2024catalyst}. Due to $\lambda_{NW,Te}$ being small, we only consider diffusion of Pb, and Sn adatoms here.   
    
    In the model it is assumed that the Group IV material can diffuse along the NW facets until it can bind with a Te adatom. Therefore, we assume that the Te-exposed facets act as a sink for the Group IV material. We assume no diffusion between facets for Te adatoms (SI 3.4).

    \item \textbf{Diffusion on the mask.} An additional contribution to the supply of material to the NW in selective-area growth is diffusion from the mask \cite{jensen2004role} \cite{Dubrovskii_2022_Adsorbing}. Based on concentration gradients, adatoms that land on the mask can diffuse over the mask and give a net material flux to the mask openings, which act as a material sink for the diffusing species. The distance these adatoms can diffuse over the mask, $\lambda_{mask}$, depends on the substrate temperature, the adatom vapor pressure and adatom interactions, and results in a collection area with a radius of $\lambda_{mask}$ around each NW. The flux of material arriving from the mask to the side facets depends on both $\lambda_{mask}$ and the pitch. For large pitches this flux is limited by $\lambda_{mask}$. For smaller pitches, collection areas of neighboring wires may overlap and competition between NWs limits the contribution from the mask. 

    As discussed above, on the mask we expect Pb, Sn, Te adatoms to be present. If only Sn is deposited on the mask, a short diffusion length, $\lambda_{mask,Sn}$, is observed(SI 3.5). However, if Te is added to the growth, an enhancement of $\lambda_{mask,Sn}$ is observed (SI 3.3). Using similar arguments as for the diffusion on the NW facets, we assume that Te itself has a short diffusion length. We hypothesize that by forming bi-atomic Group IV-Te clusters the diffusion lengths of both Sn and Te adatoms can be enhanced. We expect similar behavior for Pb adatoms. Due to the limited stability of these IV-Te pairs, their dissociation is expected during diffusion on the mask. Due to the high vapor-pressure of Te, the Te adatoms can evaporate after dissociation. However, due to the excess of Te supply on the mask, a new Te adatom can replace the previous one. This effectively increases the diffusion length of Group IV adatoms towards the NW, while the supply of Te to the NW remains limited.
    
    In the model, we account for this effect by introducing a diffusion efficiency for Te, $\eta_{diff}$. Summarizing, we identify four types of diffusing species on the mask, Pb, Sn, Te adatoms and IV-Te pairs, each with their own ‘effective’ diffusion length.

    In the growth model presented below, $\lambda_{mask, Te}$ is assumed to be 0, $\lambda_{mask, Sn}$ and $\lambda_{mask, Pb}$ are to be any positive value. The enhanced diffusion length of Sn and Pb cannot be estimated from the pitch dependency of the NW volume as is explained in SI 3.6. Contrary, by studying the pitch dependence of the NW volume, the maximum $\lambda_{mask, IV-Te}$ is determined to be 250 nm (SI 3.6). In the model, $\lambda_{mask, IV-Te}$, is between 0-250nm and $\eta_{diff}$ has to be between 0 and 1.

    \item \textbf{Shadowing.} The effect of shadowing follows directly from the presence of neighboring NWs and the assumption that the fluxes act as parallel beams  \cite{sibirev2012influence}. The impact of shadowing depends on the dimensions of the NWs and the pitch. As the sample rotates, there are different shadowing regimes (see SI 3.7). For the layout used in our experiments, shadowing contributions up-to the next-next-nearest neighbors are considered. Shadowing is a negative contribution to the flux arriving on NW side facets. This effect is expected to be important for small pitch and/or long nanowires. 
    
    Experimental evidence for shadowing is given by the pitch dependency of the NW volume of long NWs (SI 3.5). Here, the NW volume increases, with increasing pitch. We, at least partly, attribute this to a decreasing contribution of shadowing effects for increasing pitches. As the NW dimensions, the pitch, and the incidence angle of the flux are fixed input parameters for the model, shadowing is automatically assumed to be active in the model.  
 
    \item \textbf{Reflection from the mask.} It has been shown that for specific combinations of material and mask, adatoms can reflect from the mask onto a NW and contribute to crystal growth \cite{Gibson_2014} \cite{dubrovskii2022reflecting} . Usually, reflection from the mask and diffusion on the mask are mutually exclusive for a specific combination of material and mask. Specular re-emission is the most studied. The flux arriving from specular reflection to the NW side facets depends on the dimensions (collection area) of the NWs and on the pitch. As the NWs grow, the contribution increases until the shadowing of neighboring NWs starts limiting the material that hits the mask and can thus reflect on the NW. In the model, the contribution of reflection has been briefly evaluated.

    Early iterations of the model incorporated the effect of specular reflection on the NW growth. However, due to the mutual exclusivity of mask diffusion and reflection and the fact that mask diffusion was better able to explain the experimental trends, reflection is not incorporated in the final versions of the model.

    \item \textbf{Evaporation after crystal growth.} Finally, material can evaporate from the NW facets during or after crystal growth when the sample is at elevated temperatures, leading to a negative material flux \cite{dubrovskii2006theoretical}. \\To obtain an upper limit for the amount of re-evaporation at the growth conditions, a sample containing nanowires was kept at the growth temperature for an additional hour after completion of the growth, i.e. after stopping the materials fluxes. This did not result in a significant loss of NW volume (SI 3.7) compared to NWs that were cooled down directly after the growth, indicating this effect to be negligible at the growth temperatures considered in this work. 

\end{enumerate}

\subsection{Model development}

The goal of the model is to describe the evolution of the NW morphology with time. Combined with the starting geometry, the active growth factors determine the amount of Group IV and VI material present at every facet at every time step. Assuming stoichiometric growth, the growth rate for every facet can be calculated. This yields a new morphology which becomes the starting geometry of the following time step. Here, we modeled every time step of the model to be one minute, which is assumed to entail an integer number of rotations. To investigate the effect and weight of each growth factor, the model is designed in a modular fashion, such that each growth factor can be studied individually. Based on the discussion in the previous section, factors 1-5 were selected for incorporation into a numerical model. For simplicity, the model assumes an infinite array of identical NWs in a square pattern.

The nanowire dimensions at 120 minutes (from experiments) are used as starting geometry for the model to exclude effects of the nucleation phase. The other experimental input parameters are the pitch, the input flux and the incidence angle of the fluxes. 

To gain insight into the weights of the contributions of the various growth factors, we now consider these contributions individually (Figure 3). First, We consider the model with only direct impingement and shadowing (Solid blue line). Although the model portrays the super-linear behavior expected for the NW volume versus growth time (Figure 3c), the total volume is too small. This implies that at least one additional growth factor is needed to account for the missing material. Due to the chosen IV/VI ratio of 0.33 for the material fluxes, there is an excess of Te adatoms on the NW. If diffusion of Group IV adatoms from the mask is allowed (Dashed line), material is added to the NW. Here we used the arbitrary but physically reasonable diffusion lengths, $\lambda_{mask, Pb, Sn}$=400 nm, $\lambda_{mask, Te}$ =0 nm and $\lambda_{mask, IV-Te}$=250 nm. The resulting total volume corresponds much better with the experimental data set. However, the axial growth rate is still greatly underestimated (Figure 3a). Due to the supply of Group IV adatoms from diffusion from the mask, the total supply of group IV adatoms is greater than than the supply of Te adatoms from direct impingement. Thus, an excess of Group IV adatoms is created on the side facets. At the top facet, there is an excess of Te adatoms. Therefore, if we allow for facet-to-facet transfer, excess Group IV adatoms on the side facets will diffuse to the top facet and contribute to axial growth. This corresponds to the dotted lines in Figure 3. This combination of growth factors now shows the super-linear increase in volume as well as similar trends for the axial and radial growth rates. As Figure 3 shows, the best fit of the experimental data is obtained when incorporating the combination of growth factors 1-5 (direct impingement, diffusion on the NW facets, facet-to-facet transfer, diffusion from the mask and shadowing). Eliminating one or more of these factors leads to a worse fit for a wide range of input parameters SI 4. 

\begin{figure}[h]
\centering
\includegraphics[width=1\textwidth]{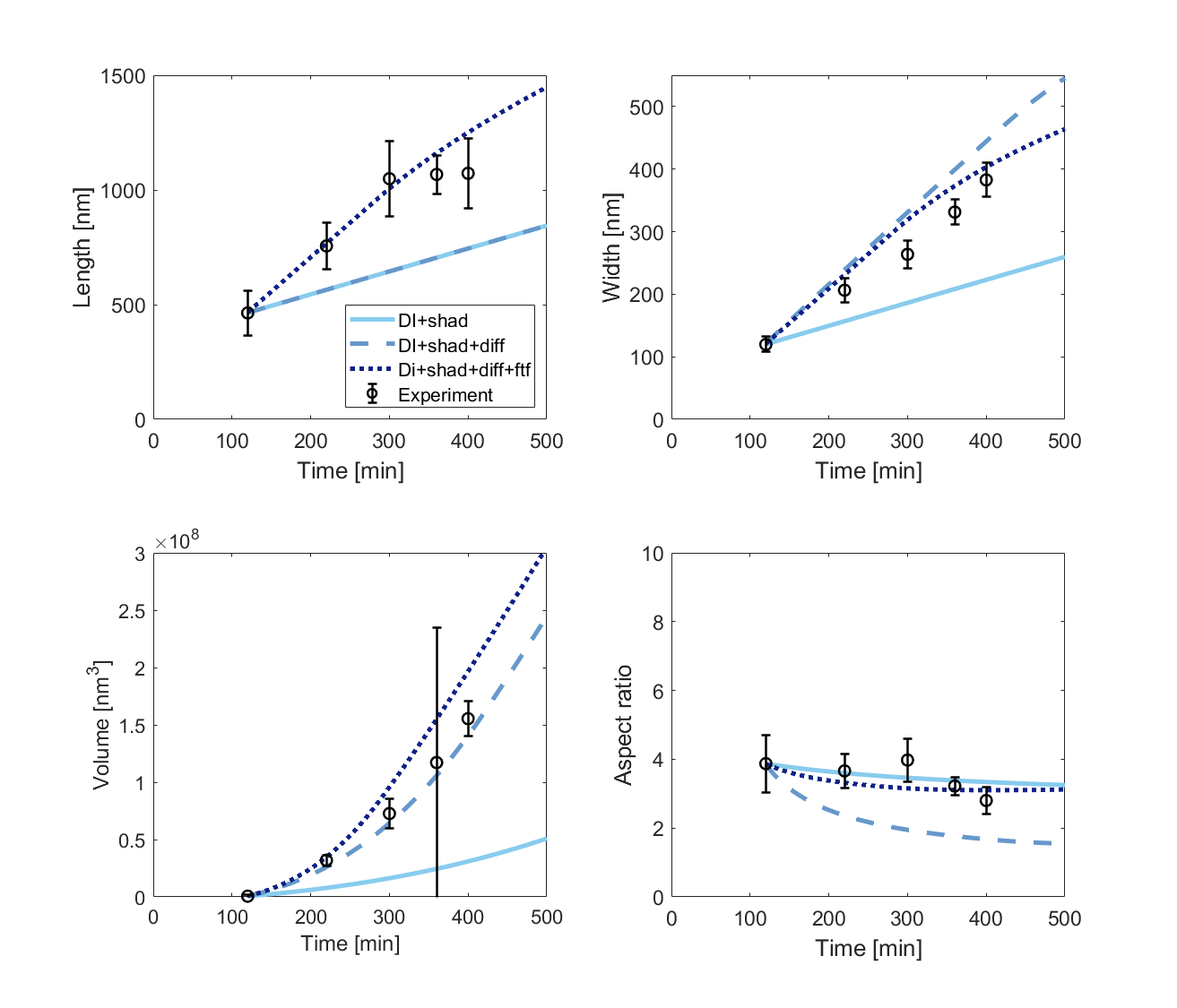}
\caption{ \fontsize{8}{14}\selectfont \textbf{Build up of the model } The NW length (a), width (b), volume (c) and aspect ratio (d) predicted by the model (lines) for varying growth factors and the corresponding experimental data (markers) for a pitch of 1000 nm and a HS of 70 nm starting at the experimental datapoint of 120 min.}
\end{figure}

In Figure 3, only one pitch and hole size are considered. In Figure 2, data and model predictions are shown for a pitch of 1000 nm and hole sizes ranging from 20 to 100 nm. The data and model predictions for the remaining pitches are presented in SI 5. We find that the model describes the experimentally observed trends well for all pitches using one set of input parameters. A good correspondence with the experimental datasets can be obtained with a range of input values for $\lambda_{mask, Sn}$, $\lambda_{mask, Pb}$ and $\lambda_{mask, IV-Te}$ , as can be seen in SI 4. The best correspondence between the model and all the experimental data is found with the following input values: $\lambda_{mask, Sn}$=350 nm, $\lambda_{mask, Pb}$=300 nm and $\lambda_{mask, IV-Te}$=250 nm. $\lambda_{mask, Sn}$ and $\lambda_{mask, Pb}$, are chosen such, that the average Sn-content is similar to the experimentally observed Sn-content as well as qualitatively matching with the Sn-gradient observed in earlier work \cite{mientjes2024catalyst}.

Now that the numerical model has proven successful in describing the growth behavior, the contribution of every growth factor per time step can be evaluated. Figures 4 a,b) show the volume contributions per time step for direct impingement and mask diffusion for a pitch of 500 nm and 2000 nm respectively. For a pitch of 2000 nm (Figure 4 b), the contribution of direct impingement increases super-linearly. This is expected as the collection area for direct impingement scales with the NW size. For a pitch of 500 nm (Figure 4a), the contribution starts super-linearly, but then flattens off and eventually decreases. This difference in behavior for small and large pitches is due to the impact of shadowing being much larger for smaller pitches.    

The behavior of the contribution of diffusion from the mask is more complex since it not only scales with the mask diffusion collection area but is also dependent on the amount of Group IV and Te adatoms at the NW facets. These contributions are shown for the NW side facets in Figure 4 c,d (solid lines) for a pitch of 500 nm and 2000 nm, respectively. At short growth times, the NW size is small compared to the collection area for mask diffusion. The amount of Te adatoms on the NW facet only scales with the NW size, since Te does not diffuse from the mask. In contrast, the collection area of Group IV adatoms on the mask is relatively large at this stage. Therefore, for short growth times, the growth, and thus the effective diffusion from the mask, is Te-limited (Figure 4 c,d). As the NWs grow, the collection area for direct impingement increases, causing an increase in the Te supply from direct impingement. Conversely, the collection area for mask diffusion gets limited by shadowing and thus the potential group IV supply from mask diffusion is limited (dashed black line). This causes a transition from a Te-limited regime to a Group IV-limited regime at 154 min and 223 min for a pitch of 500 nm and 2000 nm, respectively. This transition happens later for a pitch of 2000 nm, since there is no competition of mask diffusion collection areas for these pitches. Additionally, at a pitch of 2000 nm there is no shadowing of the mask which would otherwise limit the contribution of mask diffusion. Due to the decreasing contribution from the mask, the total volume increase per time step becomes smaller, which leads to the leveling off of the axial and radial growth rate. The final NW morphology is thus due to a balance between Te adatoms arriving from direct impingement and Group IV adatoms arriving from both direct impingement and mask diffusion.

\begin{figure}[H]
\centering
\includegraphics[width=1\textwidth]{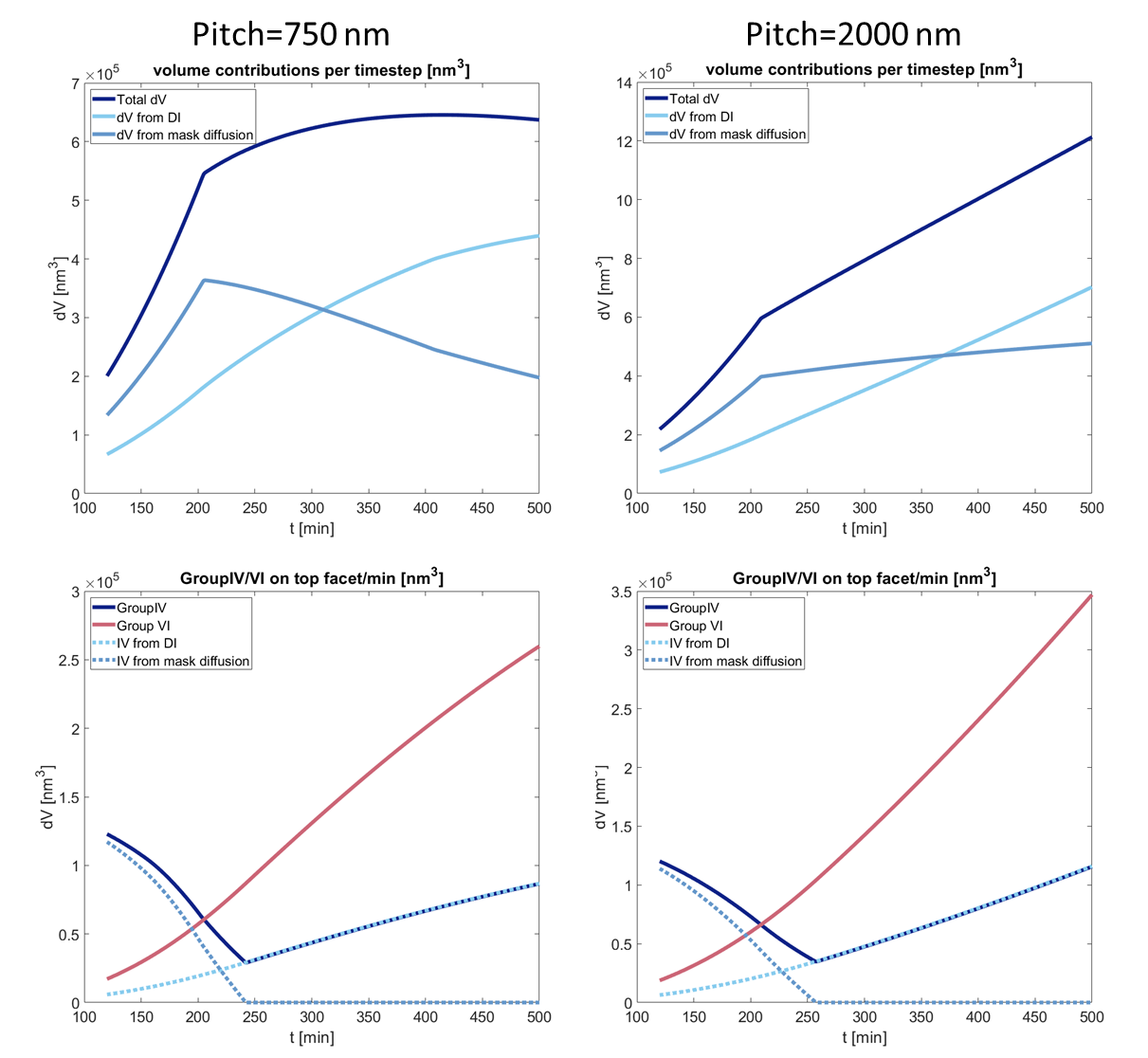}
\caption{ \fontsize{8}{14}\selectfont \textbf{Contributions of individual growth factors and Group IV/VI } a,b)The total volume contribution per time (dV) step with time and the volume contributions of direct impingement (DI) and mask diffusion individually for a pitch of 500 nm and of 2000 nm, respectively. c,d) The volume of Group IV and Te present on the side facet per time step (solid lines). The Group IV contribution is further divided into contributions from direct impingement and mask diffusion (MD) (blue dashed lines) for 500 nm and 2000 nm, respectively. The dashed black line depicts the potential Group IV material available from the mask. }
\end{figure}

\section{Conclusion}
In this work, we provided a phenomenological model to predict relevant growth factors for the VS growth of \PbSnTe nanowires. This model is consistent with an extensive series of growth experiments, and quantitatively explains the high aspect ratio of cubic nanowire growth in the [100] direction. We found that the growth is driven mainly by direct impingement and diffusion from the mask combined with the possibility for material to diffuse from one facet to another as long as there is other material to bind with. This model gives insight into the decreasing axial growth rate where the growth transitions from a Te-limited regime to a Group IV-limited regime. Although this model is specifically designed for this material system, the framework could be adapted for other material systems as the modular fashion of the model allows for easy adaptation to different material systems.

\section*{Acknowledgments}
The authors thank NanoLab@TU/e for the use of their facilities and their help and support. Additional thanks go to Sander Schellingerhout, Max Hoskam and Paulina Nowakowska for their helpful discussions. This work was supported by the European Research Council (ERC TOCINA 834290) and the Dutch government (OCENW.GROOT.2019.004).
\section*{Data availability statement}
The data that supports the findings of this study is openly available in Zenodo at DOI:10.5281/zenodo.14205503
\section*{Author contribution}
M.M. and X.G. carried out the MBE growth experiments and subsequent SEM analysis. M.M. developed the growth model. E.B. and X.G. supervised the project. M.M., X.G., M.V. and E.B. contributed to the writing of this manuscript.
\section*{Competing interests}
The authors declare no competing interests.    
   
\bibliographystyle{ieeetr}

\newpage

\end{document}